# All-fiber mode-locked ytterbium-doped fiber laser with a saturable absorber based on nonlinear Kerr beam cleanup effect

BAOFU ZHANG,[1] SHANCHAO MA,[1] SIHUA LU,[1] QIURUN HE,[1] JING GUO,[2] ZHONGXING JIAO,[1,*] AND BIAO WANG[1,3,*]

[1]School of Physics, Sun Yat-sen University, Guangzhou 510275, China
[2]School of Opto-Electronics, Bejing Institute of Technology, Beijing 100081, China
[3]Sino-French Institute of Nuclear Engineering and Technology, Sun Yat-sen University, Guangzhou 510275, China
*Corresponding author: jiaozhx@mail.sysu.edu.cn , wangbiao@mail.sysu.edu.cn

**Abstract:**
We theoretically and experimentally demonstrate a novel mode-locked ytterbium-doped fiber laser with a saturable absorber based on nonlinear Kerr beam cleanup effect. The saturable absorber was formed by a 2-m graded-index multimode fiber and a single-mode fiber segment served as a diaphragm. With an all-normal-dispersion-fiber configuration, the laser generated dissipative soliton pulses with pulse duration of 26.38 ps and pulse energy more than 0.25 nJ; output pulses could be compressed externally to 615.7 fs. Moreover, the self-starting mode-locking operation of this laser exhibited a high stability with a measured signal-to-noise ratio of 73.4 dB in the RF spectrum.

In recent years, mode-locked fiber lasers have received much attention in many commercial applications such as precise material processing [1], medical surgery [2], and spectroscopy [3]. They have become important tools in those industries and offered significant economic benefits due to their outstanding performance, compact structure, high reliability, and low cost. Utilizing fiber nonlinearities and saturable absorbers (SAs) based on special materials are two main approaches to achieve mode-locked operation in fiber lasers [4-8]. Although material-based SAs, for example semiconductor saturable absorber mirrors and various kinds of two-dimensional materials, have been used in many mode-locked fiber lasers [4, 5] and commercial products, they have relatively low damage thresholds and lifetimes which will limit the output performance of lasers. Moreover, typical mode-locked methods using fiber nonlinearities also have their shortcomings. Fiber lasers based on nonlinear polarization evolution (NPE) have low environmental stability [6]; fiber lasers using nonlinear amplifying loop mirror cannot be easily self-starting [7]. Mamyshev oscillator is a novel approach to obtain high-energy ultra-short pulses comparable to mode-locked solid-state lasers [8], however, this technique is less easily implemented in all-fiber configuration. Therefore, new methods to achieve mode-locked operation in fiber lasers are attracting widespread research interest.

Nonlinear Kerr beam cleanup effect (NL-KBC) arises when intense laser pulses with increasing power propagate in graded-index multimode fibers (GRIN MMFs) [9, 10]. During the propagation, energy of high-order modes (HOMs) is irreversibly coupled into the fundamental mode, which is mainly caused by nonlinear Kerr effect and four wave mixing in GRIN MMFs [10]. According to this feature, we have constructed SAs based on NL-KBC in our previous work [11]. Those SAs were formed by a long GRIN MMF with a short single-mode fiber (SMF) served as a diaphragm, which is easy to build up without any special material. We also experimentally investigated their characteristics and proved that they have potential to realize all-fiber configuration in mode-locked fiber laser. SAs based on NL-KBC seem to be similar to the SAs based on nonlinear multimodal interference (NL-MMI) which have been theoretically proposed [12] and experimentally applied to mode-locked lasers in different wavelength regions [13, 14]. However, unlike the SA based on NL-MMI, a seriously specific length of GRIN MMF is not required in the SA based on NL-KBC because its initial state in linear regime can be independent of the input power distribution and the length of GRIN MMF [11, 12].

In this letter, we present, to the best of our knowledge, the first all-fiber mode-locked Yb-doped fiber laser with a SA based on NL-KBC. The SA consisted of a 2-m GRIN MMF and a section of SMF served as a diaphragm; the saturable absorption measurement was performed to evaluate the properties of this SA. As the mode-locked operation self-starting, this all-normal-dispersion fiber laser could produce stable ultra-short pulses at a repetition rate of 27.57 MHz. The output dissipative soliton pulses had a pulse duration of 26.38 ps with its wavelength centered at 1043 nm. A simplified numerical simulation is also demonstrated to investigate the evolution of pulse duration and spectrum in this oscillator.

In this study, the SA based on NL-KBC was built up with a 0.4-m SMF segment (Nufern, Hi1060-XP) spliced to a 2-m GRIN MMF (Nufern GR-50/125-23HTA). When the input laser power is low, output beam profiles from the GRIN MMF and hence the transmittance of the SA can be changed by the bending condition of the GRIN MMF [10, 11]. Therefore, a pair of holders were used in the forepart of GRIN MMF to ensure the high modulation depth of the SA, and the rest of GRIN MMF was loosely coiled into a circle of about 25 cm in diameter. In order to evaluate its optical properties, we performed an in-line saturable absorption measurement on the SA [see Fig. 1(a)]. The mode-locked 1064-nm laser pulses was launched into the SA through an input SMF whose type was the same as the one used in the SA; the launched power could be adjusted by a fiber attenuator. The optical transmittance of the SA under different input intensity are presented in Fig. 1(b). The transmittance here is defined as the ratio of the output average power to the launched average power of the SA. For an SA based on fiber nonlinear effects, the transmittance curve can be fitted by

$$T(I) = T_0 + q \cdot [1 - \exp(-I/I_{sat})] \qquad (1)$$

where T is the optical transmittance, T0 is the initial transmittance, q is defined as the modulation depth, I and Isat are the input and saturation intensity, respectively. As shown in Fig. 1(b), the optical transmittance first grew monotonically as the input intensity increased and then performed a "relaxation oscillation" behavior, which has been elaborated in our previous work [11]. Moreover, the curve fits well with the experimental data; the initial transmittance, modulation depth, and saturation intensity of the SA are fitted to be 8.75%, 50.50%, and 8.17 MW/cm2, respectively. The large modulation depth and low saturation intensity of this NL-KBC-based SA are beneficial to obtain high-energy pulses with low threshold when the SA is applied to a mode-locked fiber laser.

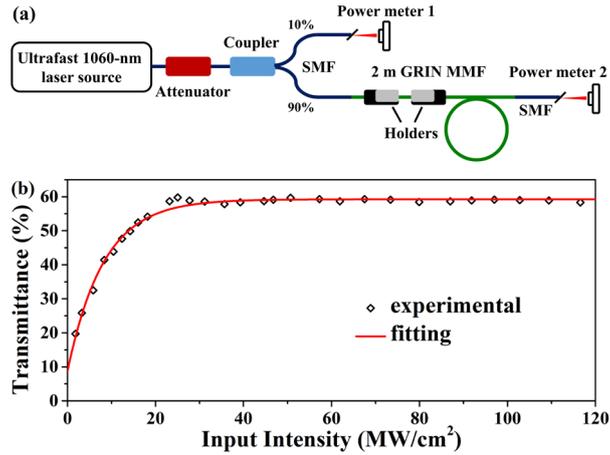

Fig. 1. (a) Experimental setup of the in-line saturable absorption measurement; SMF: single-mode fiber, GRIN MMF: graded-index multimode fiber. (b) The optical transmittance of the saturable absorber with increasing input intensity.

The experimental setup of our mode-locked laser with an all- fiber ring cavity configuration is depicted in Fig. 2. A 976-nm single- mode laser diode was used as the pump source. Its pump power was launched into an ytterbium-doped fiber (Yb1200-4/125, LIEKKI) through a wavelength division multiplexer. A coupler was located after the gain fiber, and its output ratio of 10% was designed for low mode-locked threshold. The output port of the coupler was spliced to an isolator which can protect the stable mode-locked operation from the interference of reflected beams. A polarization independent isolator (PI-ISO) was also used in the cavity to ensure single direction operation and no other mode-locking mechanisms (especially NPE) in the laser. The SA based on NL-KBC was the same as what we mentioned above. Two polarization controllers were placed before and after the SA; they were used for not only ensuring the linear polarization state of the input beam to achieve NL-KBC effect in the SA but also optimizing the intra-cavity birefringence to obtain stable mode-locked operation of the laser. All single-mode passive fibers used in the cavity had the same type as the one used in the SA (Nufern, Hi1060-XP), and hence the laser had an all-normal-dispersion-fiber cavity. Moreover, the band-pass filtering effect were produced by the PI-ISO and the SA [15], respectively, which is beneficial to achieve dissipative soliton output from this all-normal-dispersion cavity. The linear transmission spectra of the PI-ISO and the SA were measured by using a broadband light source with its output spectrum ranged from 1000 to 1130 nm. The central wavelength and 3 dB bandwidth of PI-ISO were measured to be 1044 nm and 8.5 nm, respectively; the 3 dB bandwidth of SA was measured to be about 50 nm and its central wavelength could be changed with the bending condition of GRIN MMF.

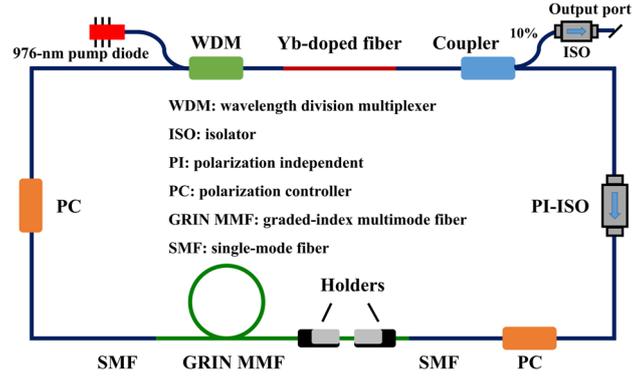

Fig. 2. Experimental setup of the mode-locked ytterbium-doped fiber laser with a saturable absorber based on nonlinear Kerr beam cleanup effect.

Firstly, we investigated the evolution of pulse and spectrum in this oscillator by numerical simulations based on split-step Fourier method [16]. The Kerr nonlinearity, the second-order, and third-order dispersion are included in the simulation. Since the complicated mechanism of NL-KBC effect [10] cannot be easily embedded in this simulation, we consider that only fundamental-mode beam is propagating in the GRIN MMF. Therefore, the GRIN MMF segment in our model is simplified to be a 2-m SMF with the same fundamental-mode diameter, a virtual SA, and a virtual band- pass filter with their parameters as measured above at the end of this 2-m SMF. Moreover, a simplified uniform gain dynamic in the ytterbium-doped fiber with the assumed Lorentzian bandwidth of 30 nm centered at 1038 nm, and other fiber components with their accurate parameters are employed in this model. A stable mode- locked regime in which the laser produces 0.20-nJ output pulses can be found, its simulated results are shown in Fig. 3. In this case, although the oscillator can be seeded with different initial pulses, the simulations with given cavity parameters will converge to the same solution in a few round trip. As shown in Fig. 3(a), the laser pulse duration and spectral bandwidth grow monotonically in the passive fiber, but the pulse duration slightly decreases while the spectral bandwidth continues to grow in the gain fiber. The laser pulse is shaped by the SA and band-pass filters to a profile with shorter duration and narrower bandwidth which could seed the oscillator in the next round trip. Spectral breathing by a factor of only 1.15 can be observed due to the relative low gain of the operating wavelength and weak band-pass filtering effects. As shown in Fig. 3(b), parabolic shaping effects in the time domain cannot be found in the gain fiber segment, which is different from the self-similar regimes [17]. Moreover, the numerical results with a non-parabolic output pulse profile in the time domain [Fig. 2(c)] and a cat-ear-like output spectrum [Fig. 2(d)] have further confirmed that the laser has generated dissipative solitons.

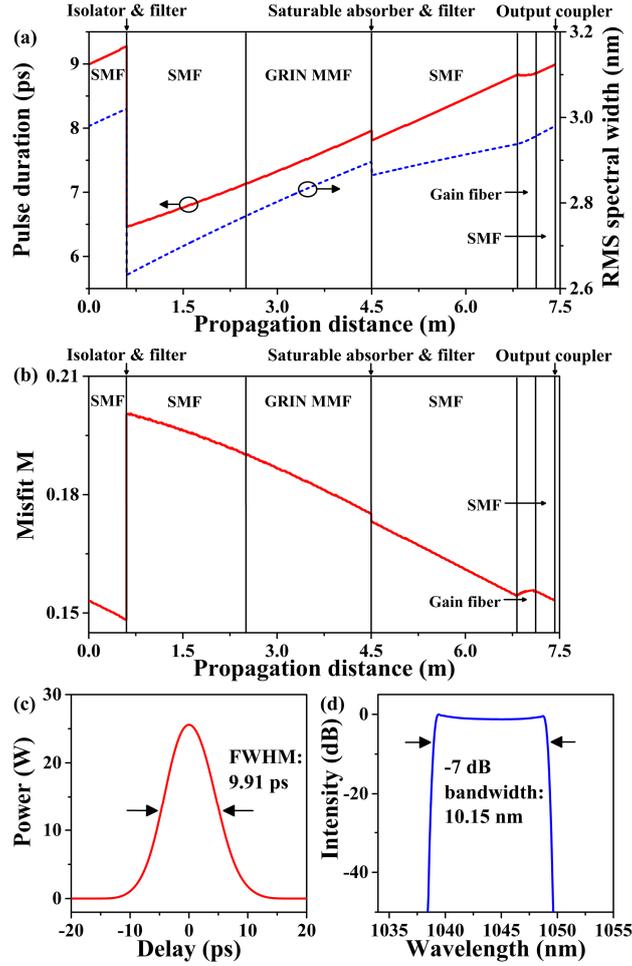

Fig. 3. The numerical results of the laser: (a) the evolution of pulse duration and spectral width in the cavity; (b) the evolution of parabolic misfit parameter M defined by $M^2= \int (I-I_{fit})^2 dt / \int I^2 dt$ in the cavity; (c) the simulated output pulse profile; (d) the simulated output spectrum profile. The black solid lines in (a) and (b) are used to indicate different regions. SMF: single-mode fiber, GRIN MMF: graded-index multimode fiber, RMS: root-mean-square, FWHM: full width at half maximum.

With the guidance of simulations, the length of ytterbium-doped fiber was designed to be 0.3 m for sufficient absorption of the pump power; the total cavity length was designed to be 7.4 m. The self-starting mode-locked operation of the laser was observed when the input pump power was increased to 60 mW. When the input pump power was up to 170 mW, the laser could still maintain a typical fundamental-mode-locking state as shown in Fig. 4(a). Here, the laser produced stable pulse train with a small pulse-to-pulse intensity fluctuation of 0.70%, and its output power was measured to be 7.02 mW. When the pump power reached a higher level, the laser exhibited a complex multiple-pulse operation dynamic. Figure 4(b) illustrates the output RF spectrum of the laser when the input power was 170 mW. The RF spectral center was measured to be 27.57 MHz, which corresponds to the ring cavity length of 7.4 m. The signal-to-noise ratio of the RF spectrum was as high as 73.4 dB; no signal of multi-pulsing or harmonic mode-locking could be found in the long-span RF spectrum. These results further confirm stable fundamental-mode-locking operation of the laser.

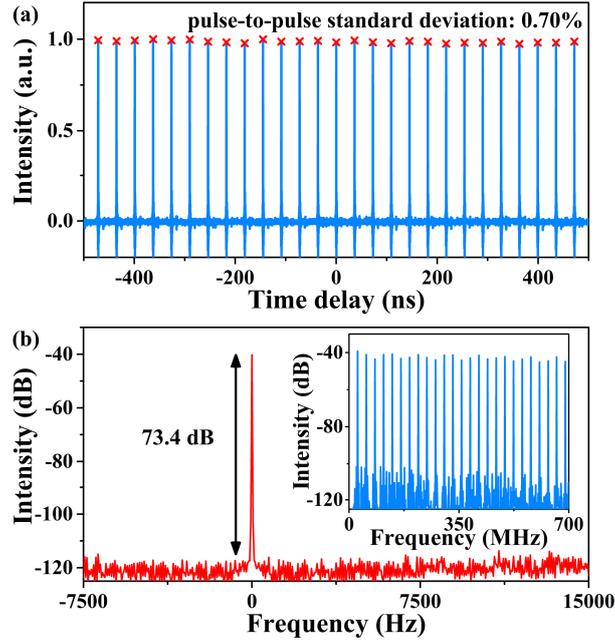

Fig. 4. (a) The output fundamental-mode-locking pulse train. (b) The output radio frequency spectrum with a resolution bandwidth of 1 Hz and a span range of 30 kHz; the central frequency is shifted to zero for clarity. Inset: the output radio frequency spectrum with a resolution bandwidth of 100 Hz and a span range of 700 MHz.

Besides, we have measured the output optical spectrum and the autocorrelation trace of output pulses. As shown in Fig. 5(a), the optical spectrum presented a cat-ear-like trace which is a typical behavior of dissipative solitons. The spectrum was centered at 1043.4 nm due to the band-pass filtering effect of the PI-ISO, and its -7 dB bandwidth was about 9.2 nm. The spectrum profile was slightly asymmetric since the operating wavelength is away from the center of gain spectrum, which was in good agreement with the simulated result. No other operating wavelengths can be observed in the long-span spectrum, and the output pulse energy was calculated to be 0.254 nJ. The autocorrelation trace of the output pulse is shown in the inset of Fig. 5(b). Assuming a Gaussian shape, its FWHM was measured to be 37.30 ps, and hence the pulse width was calculated to be 26.38 ps. The differences in chirped states and pulse widths between the experimental and simulated results might be caused by the simplified model of GRIN MMF segment in the simulation. In fact, the modal dispersion and high-order nonlinearities in the GRIN MMF can play an important role in the pulse evolution, but they were not included in this simplified numerical model. Except the chirped state of the pulse, the simulations fitted well with the experimental results. Moreover, the output pulses were dechirped externally using a homemade compressor with a single 1200 line/mm diffraction grating. The minimum pulse duration of 615.7 fs was obtained as shown in Fig. 5(b).
Although the mode-locking operation of this fiber laser with NL- KBC-based SA could maintain for more than 6 hours in a lab environment, it was not environmentally stable in the conventional sense because all fibers used in the cavity were not polarization maintaining (PM). As a result, a small disturbance to the fiber would lead to the destabilization of the output pulses. Moreover, since a pair of holders were used to manage the modulation depth of the SA, the mode-locking threshold and output pulse performance would be slightly different if the bending condition of the GRIN MMF was changed. For example, with different bending conditions of the GRIN MMF, the fundamental-mode-locking threshold could range from 60 mW to 170 mW in our case.

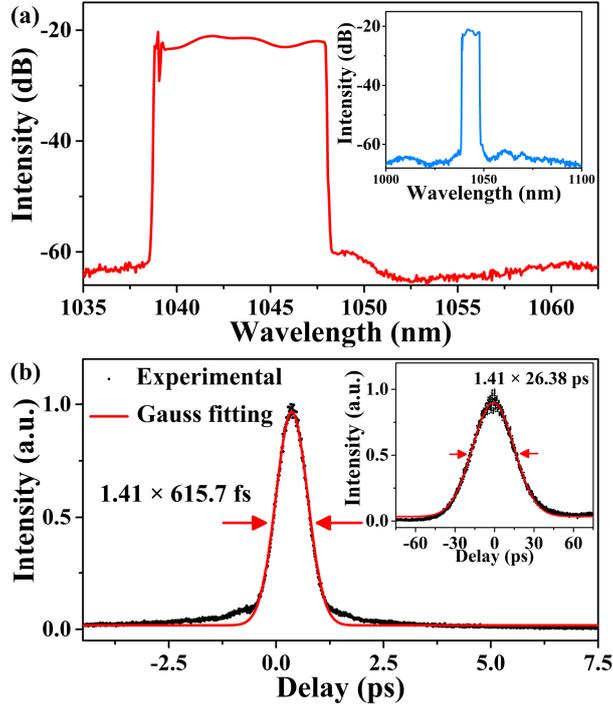

Fig. 5. (a) The output optical spectrum with a spectral resolution of 0.06 nm and a span range of 30 nm. Inset: the output optical spectrum with a spectral resolution of 0.2 nm and a span range of 100 nm. (b) The autocorrelation trace of dechirped output pulses with Gaussian fitting. Inset: the autocorrelation trace of output pulses with Gaussian fitting.

In conclusion, we first demonstrate a mode-locked ytterbium- doped fiber laser with a NL-KBC- based SA. The SA only consisted of a 2-m commercial GRIN MMF spliced with a SMF segment, and thus the laser cavity had a compact and all-normal-dispersion-fiber configuration without any special fibers or devices. With the guidance of numerical simulations, the stable fundamental-mode-locking operation of the laser could be self-starting with a signal-to-noise ratio of 73.4 dB in the RF spectrum. The laser produced 26.38-ps dissipative soliton pulses at 27.57 MHz with its pulse energy more than 0.25 nJ, and its output pulses could be dechirped to 615.7 fs. We think the output energy can be further increased with larger pump power, optimized gain fiber length and output coupling radio. Moreover, according to the behaviors of NL-KBC effect, this mode-locking method with the NL-KBC-based SA can be easily applied to the lasers in different wavelength regimes. Future works should pay more attention to the PM GRIN MMFs, which can be the key point to achieve environmentally stable operation of these lasers.

**Funding**. National Natural Science Foundation of China (11832019, 11472313, 13572355); Natural Science Foundation of Guangdong Province (2017A030310305, 2018A030310092).

**Acknowledgment**. The authors would like to acknowledge Yu Duan of Emgo-tech Ltd (Zhuhai), Fujuan Wang, and Jiaoyang Li for the technical support in this work.

**Disclosures**. The authors declare no conflicts of interest.

References
Set references at the back of the manuscript. Optics Letters uses an abbreviated reference style. Citations to journal articles should omit the article title and final page number. However, full references (to aid the editor and reviewers) must be included as well on a fifth page that will not count against page length.